\documentclass[%
 reprint,
nofootinbib,
 amsmath,amssymb,
 aps,
]{revtex4-2}

\usepackage{graphicx}
\usepackage{dcolumn}
\usepackage{bm}
\usepackage{xcolor}
\usepackage{makecell}

\begin{document}

\preprint{APS/123-QED}

\title{Atmospheric Muon Measurements Near Tornadic and Non-Tornadic Storms in the US Central Plains}

\author{William Luszczak}
\affiliation{
Department of Astronomy, Ohio State University\\
}%
\affiliation{
Department of Physics and Center for Cosmology and Astroparticle Physics, Ohio State University
}%

\author{Jana Houser}
\affiliation{
Department of Geography, Ohio State University
}%

\author{Matt Kauer}
\affiliation{
Department of Physics, University of Wisconsin-Madison\\
}%
\affiliation{
Wisconsin IceCube Particle Astrophysics Center, University of Wisconsin-Madison
}%
\author{Leigh Orf}
\affiliation{
Cooperative Institute for Meteorological Satellite Studies, University of Wisconsin-Madison\\
}%

\date{\today}

\begin{abstract}
    Tornadoes and other severe weather hazards affect thousands of people every year. Despite this, the details surrounding tornadic processes including formation, decay, and longevity are not well understood, partially due to limitations of available instrumentation. Measurements of atmospheric pressure within tornadic systems currently rely almost entirely on in-situ instrumentation, and no existing techniques can provide two-dimensional spatial information of the atmospheric density field. Atmospheric muons may hold a solution to this problem: muons are attenuated by matter, and tornadic storms are large regions of low atmospheric density, suggesting that tornadic storms induce a directional perturbation on the atmospheric muon flux. Measurements of this perturbation could then be used to infer the density field associated with severe weather. Simulations of these systems indicate that a robust measurement of the atmospheric density field would require a relatively large muon detector, however smaller detectors may be able to detect ambient muon flux perturbations if the storm is large and intense enough. This paper presents results from a pilot field study that measured the atmospheric muon flux near tornadic storms during May 2025, including directional measurements of the muon flux near tornadic mesocyclones and a measurement of the muon flux near the base of a forming tornado.
\end{abstract}

\maketitle

\section{Introduction}
\subsection{Mesocyclones and Tornadoes}
Tornadoes are violently rotating columns of air that extend from the ground to the base of a thunderstorm. While tornadoes can occur in many locations throughout the world, they are particularly common in the United States Great Plains. Despite their ubiquity, the processes by which tornadoes form are not entirely understood. It is known that the strongest tornadoes form from supercell thunderstorms, which are defined by having persistent rotating updrafts called mesocyclones. Both the tornado and parent mesocyclone are known to induce local air pressure perturbations over large volumes, with the most intense deviations found within the tornado itself.

While theory and simulation efforts have indicated the presence of pressure perturbations within tornadic storms~\cite{AStudyoftheTornadicRegionwithinaSupercellThunderstorm,BuoyancyandPressurePerturbationsDerivedfromDualDopplerRadarObservationsofthePlanetaryBoundaryLayerApplicationsforMatchingModelswithObservations,PerturbationPressureFieldsMeasuredbyAircraftaroundtheCloudBaseUpdraftofDeepConvectiveClouds}, the experimental measurement of this effect has proven difficult, and density measurements are absent altogether. Previous measurements of the ground-level air pressure within tornadic storms have relied on deployment on surface-based in-situ instrumentation in the direct path of the storm, and elevated measurements of pressure have only been acquired peripherally around the storm using instrumented aircraft. Given the significant logistical challenges involved, such measurements are understandably rare. Additionally, existing instruments that measure pressure and temperature rely on point-based direct observations  only capable of measuring atmospheric values at the sensor location. This restriction limits the amount of information available even with a successful deployment. Furthermore, no observations of pressure have been successfully collected from within a supercell storm owing to the challenges of navigating an aircraft within the incredibly hostile environment of the supercell storm, considering its large vertical velocity gradients and the threat of very large hail. The only methodologies that exist to derive estimate pressure fields are from derivations based upon radar multi-Doppler syntheses~\cite{BuoyancyandPressurePerturbationsDerivedfromDualDopplerRadarObservationsofthePlanetaryBoundaryLayerApplicationsforMatchingModelswithObservations}, or from numerical simulations of supercells~\cite{AStudyoftheTornadicRegionwithinaSupercellThunderstorm}. 

Though existing measurements and field studies have made great strides in improving scientific understanding of tornadic storms, the lack of robust observational data sets describing the atmospheric parameters within storms that form tornadoes and likewise within storms that do not, for comparison, is a significant roadblock to further progress. While novel observational techniques that expand our ability to measure the atmospheric state near tornadic storms will not single-handedly allow for perfect forecasting of severe weather systems, they will undoubtedly play a role in significantly advancing scientific understanding of tornadogenesis and the processes associated with storm evolution that produce them.

\subsection{Cosmic Rays and Atmospheric Muons}
Cosmic rays are charged particles (protons, as well as heavier atomic nuclei) accelerated in a variety of astrophysical sources, with the majority of low-energy cosmic rays originating from the Sun. Cosmic rays that reach Earth can interact with nuclei in the atmosphere, producing a shower of secondary particles. Among these particles are muons, produced primarily via pion and kaon decay:

\begin{equation}
\pi^{\pm} \rightarrow \mu^{\pm} + \overset{\textbf{\fontsize{2pt}{2pt}\selectfont(---)}}{\nu}_{\mu}
\end{equation}
\begin{equation}
K^{\pm} \rightarrow \mu^{\pm} + \overset{\textbf{\fontsize{2pt}{2pt}\selectfont(---)}}{\nu}_{\mu}
\end{equation}

The energy loss of the resultant muon is proportional to the density of matter traversed~\cite{RadiationTextbook}, resulting in muons traveling shorter distances through matter of higher density. A consequence of this is the proportionality of the atmospheric muon flux (measured in units of (GeV cm$^2$ s sr)$^{-1}$) to the local atmospheric pressure and temperature~\cite{Jourde_2016, tilav2019seasonal}.

Since atmospheric muons (mostly) travel in straight lines, and are only affected by the portion of the atmosphere they individually traverse, directional maps of the atmospheric muon flux intensity can be used to image density perturbations in the atmosphere. From the perspective of a muon detector, directions with lower integrated atmospheric density will result in a higher muon flux, while directions with a higher integrated atmospheric density will result in a lower muon flux.

The above description is an extension of a technique (``muography") used to map the interior of the Great Pyramids~\cite{LECHMANN2021103842} and volcanoes~\cite{PMID:31040358}. In recent years atmospheric muography has been used to image typhoons~\cite{typhoons} and non-tornadic thunderstorms~\cite{muthunderstorms2}. Owing to the relationship between pressure and density through the ideal gas law, it is reasonable to expect variations of pressure within storms will also be associated with simultaneous variations in density of equal sign. For example, the rotating mesocyclone is typically associated with a low pressure perturbation. If the temperature between the mesocyclone and its surroundings is assumed constant, this would imply lower density in the mesocyclone, in the region of the low pressure perturbation. In some cases, there is evidence to suggest the mesocyclone is associated with both a warm temperature perturbation and a low pressure perturbation, which would imply an even more notable decrease in density than if the temperature were constant. 

Simulations of tornadic mesocyclones suggest that the associated atmospheric density perturbation may be detectable with a large enough detector~\cite{PhysRevD.111.023018}, though there are multiple logistical and engineering challenges associated with transporting a muon detector near severe weather systems. This study aims to explore the logistical feasibility of measuring atmospheric muon flux variations near tornadic storms by using a small ($\approx$1 m$^2$) muon detector to constrain the scale of muon flux variations near severe weather systems in the United States Great Plains. The measured muon flux can then be used in tandem with available radar data to infer bounds on mesocyclone size and associated density perturbation, a quantity that is currently largely unconstrained for tornadic mesocyclones.   

\subsection{Atmospheric Density Measurement with Atmospheric Muons}
\begin{figure}[h]
    \centering
    \includegraphics[width=0.45\textwidth]{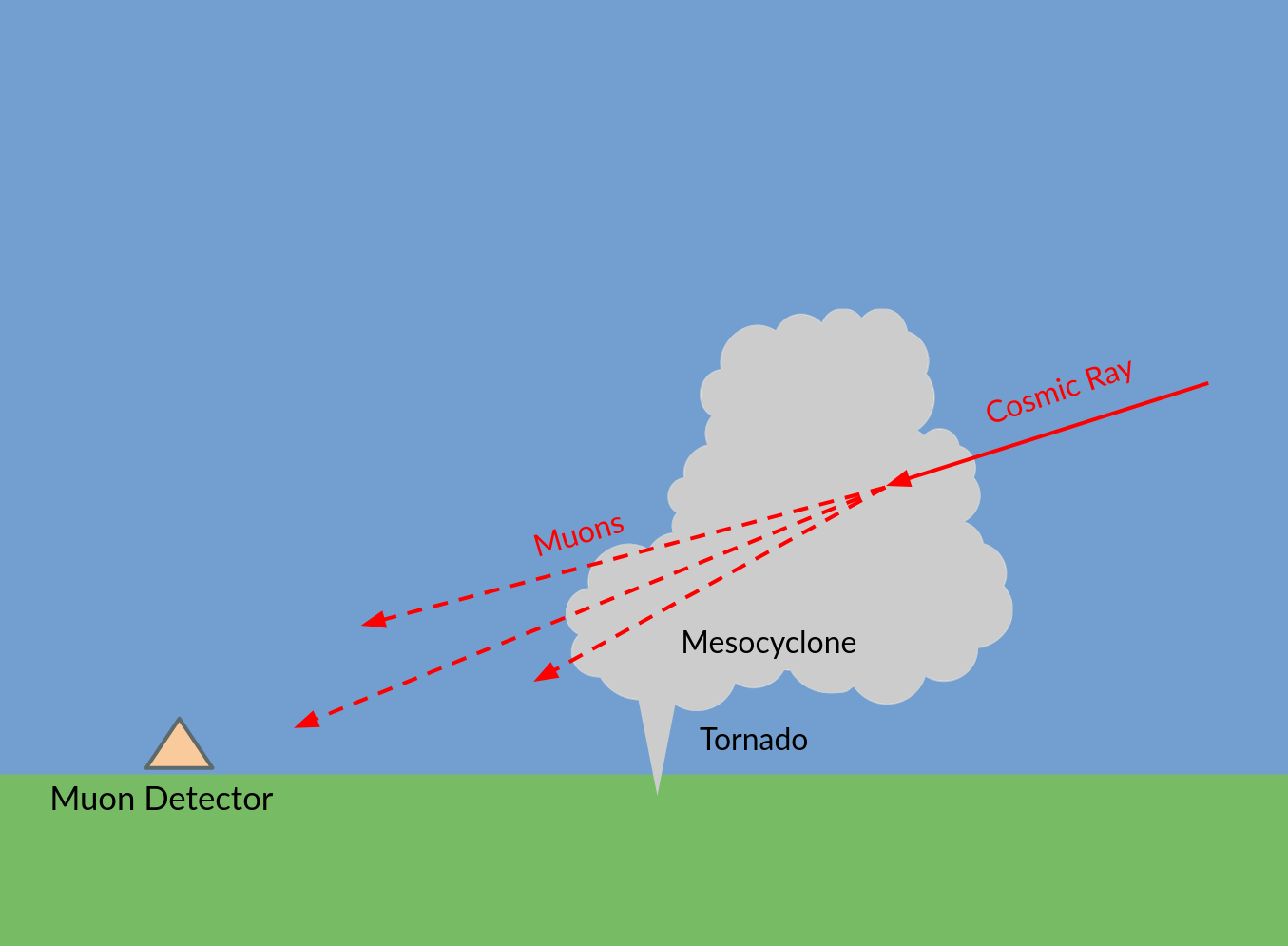}
    \includegraphics[width=0.45\textwidth]{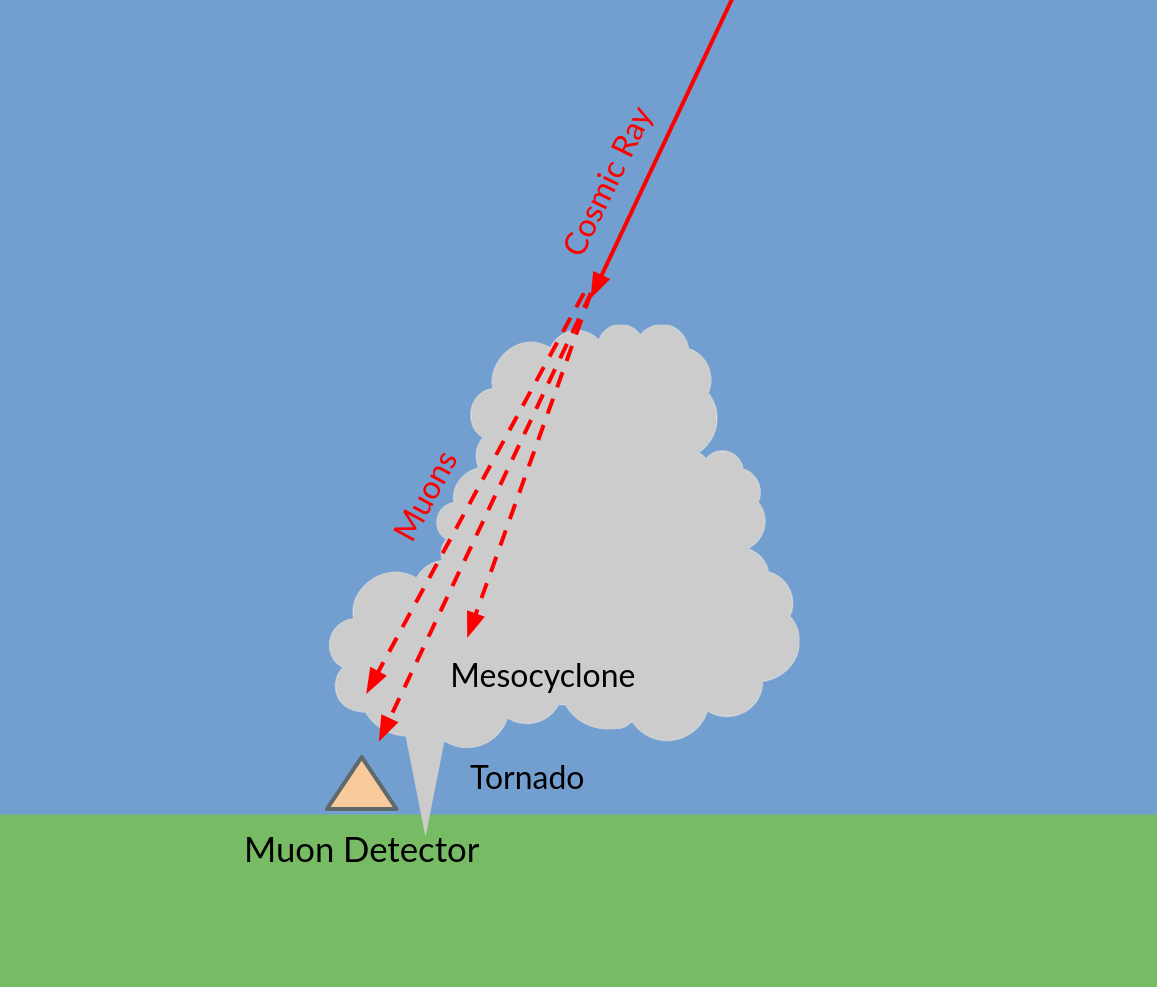}
    \caption{Top: A cartoon showing the ideal setup for this study. A muon detector is placed some distance away from a mesocyclone of interest, and the atmospheric muon flux from that direction is measured. The muon flux from the opposite direction (ideally without any severe weather) can then be used as a control measurement. Bottom: An alternative setup in which the muon detector is placed close to the severe weather system of interest. In contrast with the ideal setup, this setup boasts improved sensitivity to local density perturbations, however separate control measurements must be obtained after the detector has been moved away from the severe weather system.}
    \label{fig:setupcartoon}
\end{figure}

The generic setup of atmospheric density measurement with atmospheric muons is fairly straightforward in theory. A muon detector merely needs to be placed near a severe storm of interest, as shown in figure~\ref{fig:setupcartoon}. This study uses a muon detector that can independently measure the muon flux in two opposite directions. One side of the muon detector is faced towards the storm system of interest, while one is faced away. The muon flux rates between the two directions can then be compared, establishing a difference in integrated atmospheric density between the two directions. 

As the muon flux falls as zenith angle increases~\cite{GaisserCR}, there is some optimization to be made with determining exactly how far away to place the muon detector. If the detector is placed too far away from the storm, the muon flux from the direction traversing the storm is low, and the measured muon flux difference between the storm and clear air may be too small to measure over reasonable observation periods. A muon detector placed too close to the storm observes more muons traversing the storm, but must contend with potentially damaging winds, and loses many of the advantages associated with being a remote measurement.

Previous studies of both supercellular structure  and simulations of muographic measurements near mesocyclones indicate that the density perturbation associated with the mesocyclone may be up to 10 kilometers tall, and that the optimal viewing elevation angle for muographic measurements of these systems is at least $45^{\circ}$~\cite{Orf2019-kn, PhysRevD.111.023018}. This viewing angle maximizes distance from the storm while maintaining a large enough directional muon flux to be sensitive to realistic density perturbations associated with a mesocyclone. For the purposes of this study, the deployment team attempted to place the muon detector within 10 kilometers of the target weather systems, with varying degrees of success.

An alternative setup can be seen in the bottom panel of figure~\ref{fig:setupcartoon}. In this arrangement, the muon detector is placed much closer to the storm of interest. This improves the sensitivity of the technique, as the relevant part of the storm is at a high elevation angle, allowing for the observation of more muons per unit time from this direction. This in turn results in a more precise measurement of the muon flux over a fixed time interval, corresponding to sensitivity to smaller atmospheric density perturbations. The drawbacks of this setup are the logistical challenges involved in deploying a muon detector close to the storm, and the need for a separate control measurement after the storm is no longer near the muon detector.

\section{Instrument Description}

\subsection{Overview}
\begin{figure}[h]
    \centering
    \includegraphics[width=0.45\textwidth]{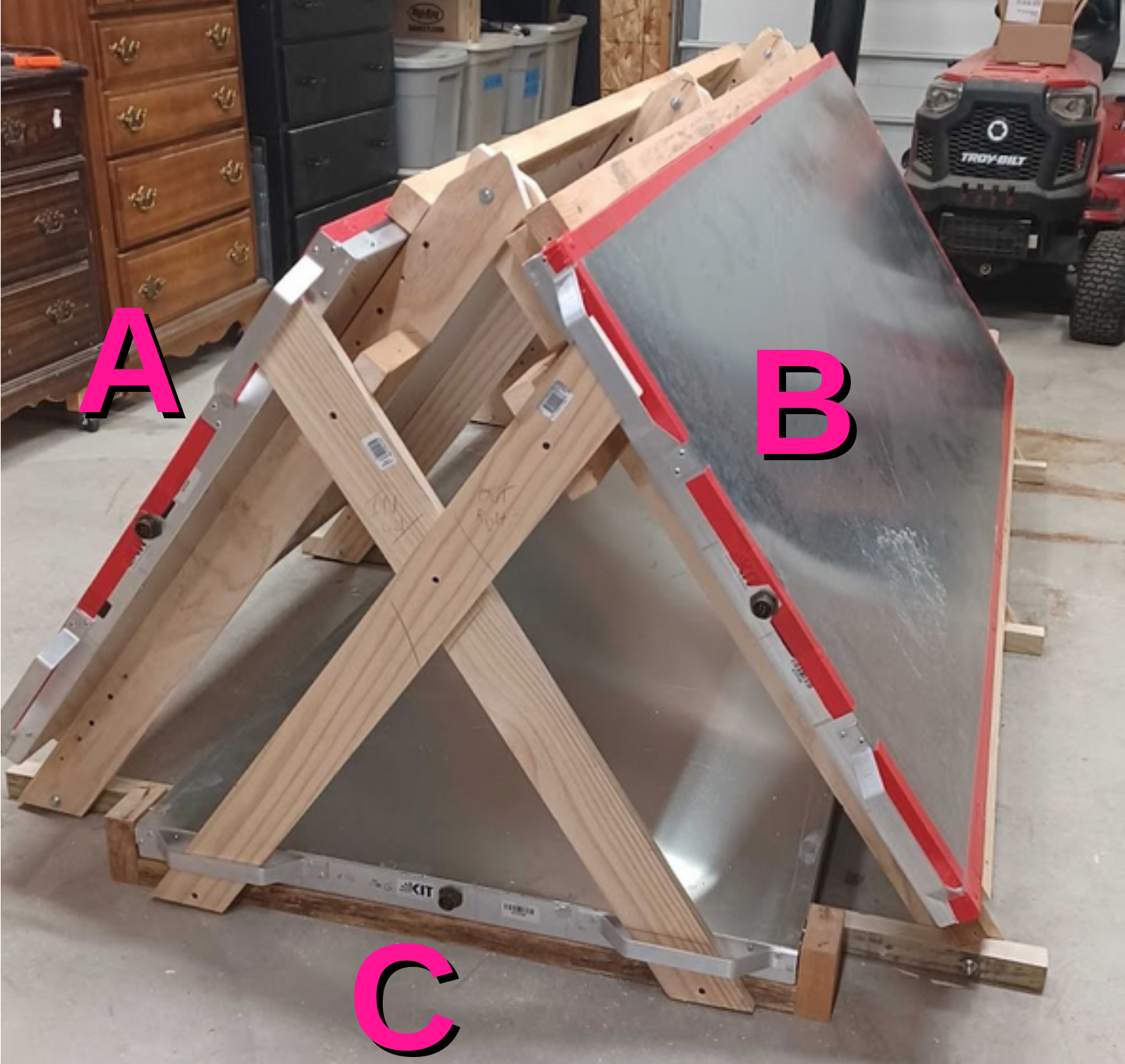}
    \includegraphics[width=0.45\textwidth]{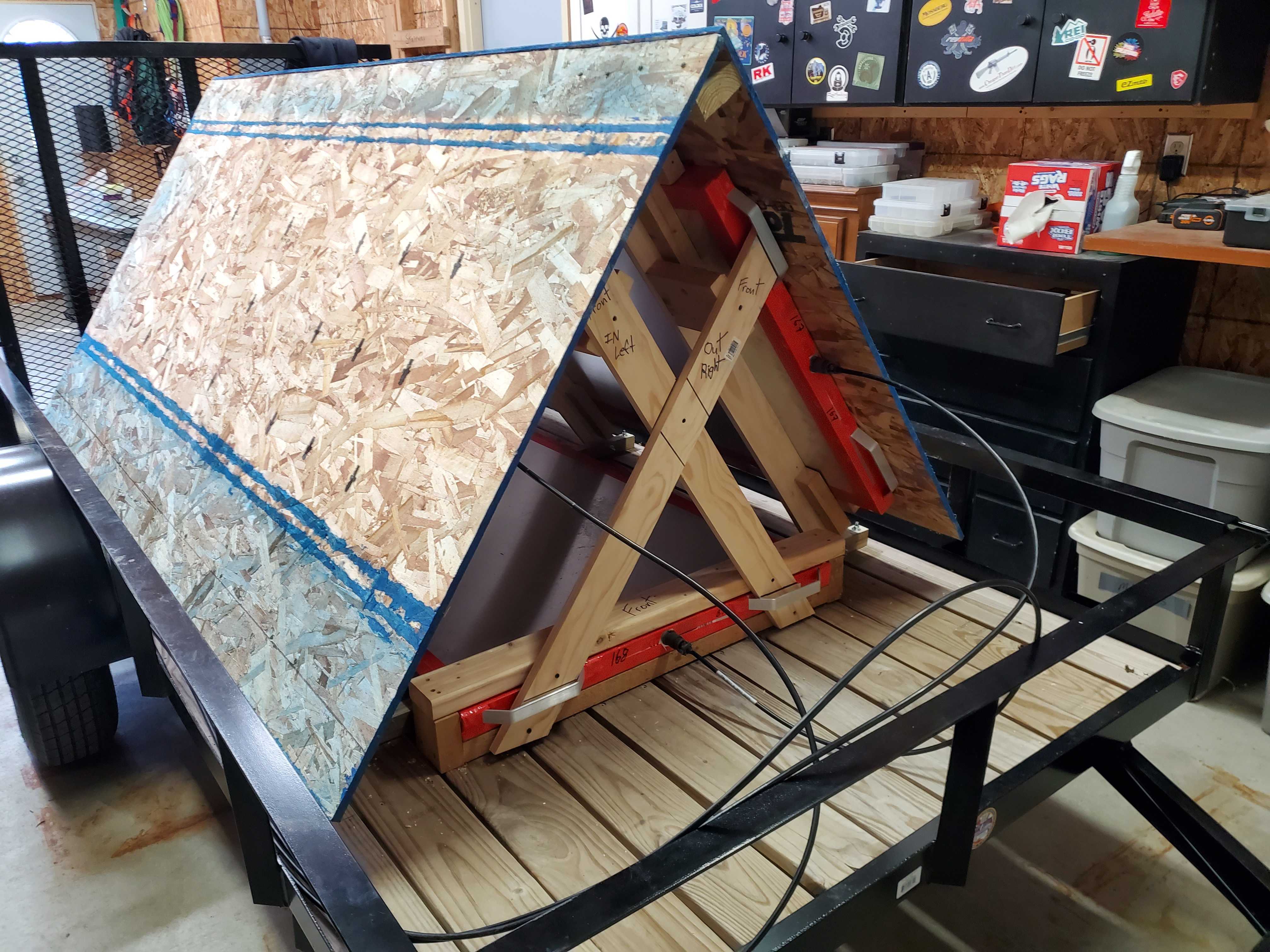}
    \caption{Photos of the muon detector used in this project, showing the three detection planes (denoted A, B, and C) mounted into a triangular configuration (top). Muon direction can be roughly reconstructed by examining coincidences between pairs of panels. A coincident detection between A and C corresponds to a muon originating from the left in the figure, and a coincidence between B and C corresponds to a muon originating from the right. Coincidences between A and B are not recorded.The entire device was mounted on a trailer and covered with protective plywood (bottom).}
    \label{fig:detpics}
\end{figure}

The muon detector used in this project is a bi-directional, coincidence detector constructed from three plastic scintillator detection planes arranged into a triangle, as shown in figure~\ref{fig:detpics}. The geometry of the device allows for rough reconstruction of individual muon event direction based off of which pairs of panels observed a coincident signal. 

\begin{figure}[h]
    \centering
    \includegraphics[width=0.45\textwidth]{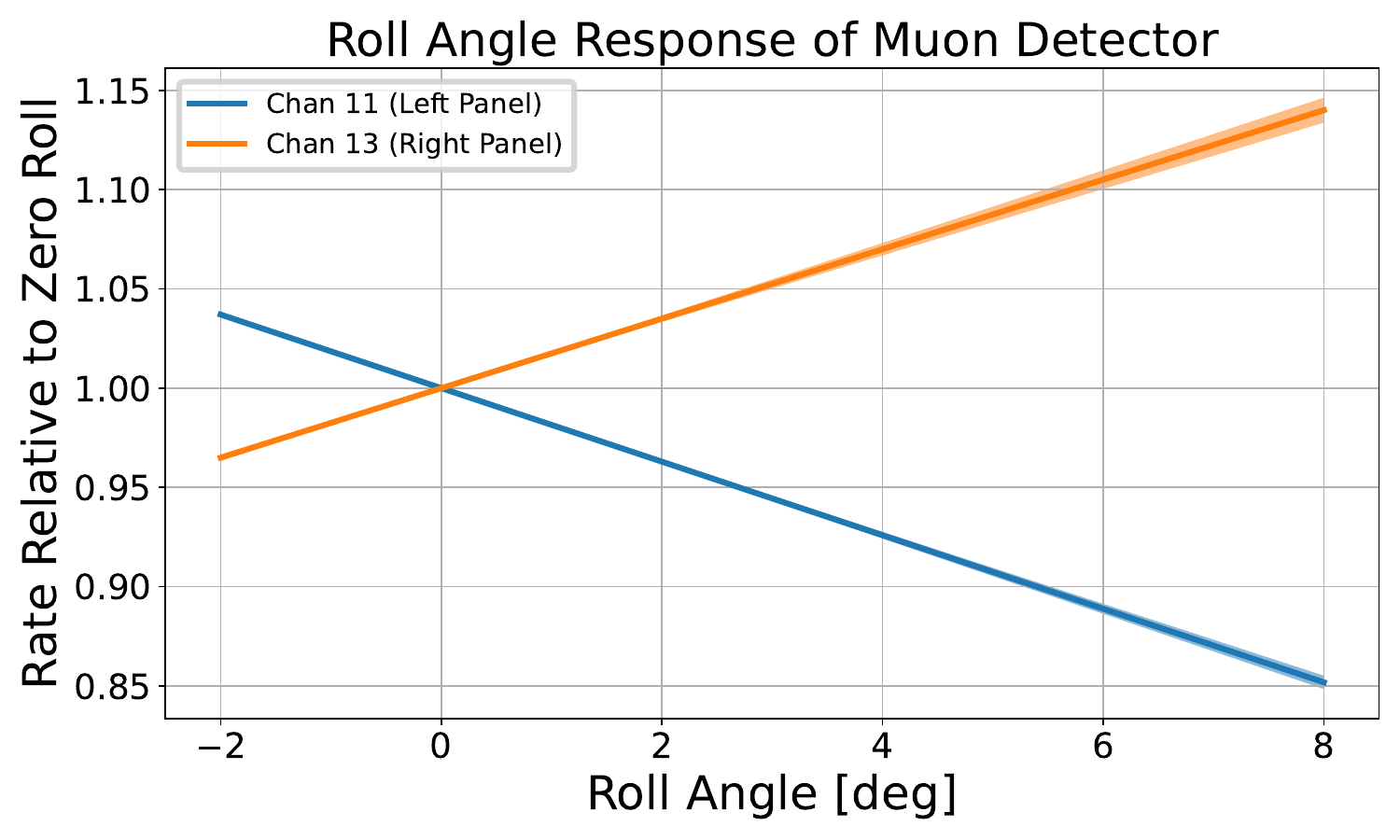}
    \caption{Calibration measurements used to account for detector roll angle when observing storms. ``Chan 11" corresponds to the rate observed by the pair of panels B and C, while ``Chan 13" corresponds to the rate observed by the pair of panels A and C.  }
    \label{fig:rollangle}
\end{figure}

The relative rates measured by each pair of panels (A and C vs. B and C) are affected by the detector roll angle, with the more vertical pair of panels measuring a larger rate of muons. This effect is corrected for with measurements of the detector roll during deployment in combination with calibration measurements of detector roll angle obtained post-deployment on clear-sky days. The dependence of the relative muon rates measured by each pair of panels on detector roll angle can be seen in figure \ref{fig:rollangle}.

Each scintillation detector weighs less than 50 kg and has an active area of 1.5\,m$^2$. It consists of 16 plastic scintillation bars with a width of 5\,cm and a reflective coating of TiO$_2$ produced by FNAS-NICADD~\cite{Beznosko:2005ba}. When an ionizing particle crosses the active area, UV light is produced. Wavelength-shifting fibers (Y-11(300) by Kuraray) are routed through two holes in the scintillation bars to guide the light produced in the bar to a Silicon Photomultiplier (SiPM) (S13360-6025PE by Hamamatsu). Since SiPMs have a temperature dependent behavior, a temperature sensor, placed close to the SiPM, monitors the temperature during operation. Opaque electro-static discharge (ESD) foil is used for light shielding. The entire scintillation detector is contained in an aluminum, weatherproof housing.

The design and construction of these panels is identical to those planned for the IceTop detector enhancement~\cite{Kauer:2019len}. Muons passing through the plastic scintillator produce light that is then detected by the SiPM, converting this light into a detectable $\sim$100\,ns duration voltage pulse that can be amplified and forwarded to triggering electronics. 

The entire arrangement of 3 panels is covered in plywood and mounted on a trailer, allowing for rapid transportation to weather systems of interest. The completed setup can be seen in figure~\ref{fig:detpics}.

\section{Observations}
\begin{figure}[h]
    \centering
    \includegraphics[width=0.45\textwidth]{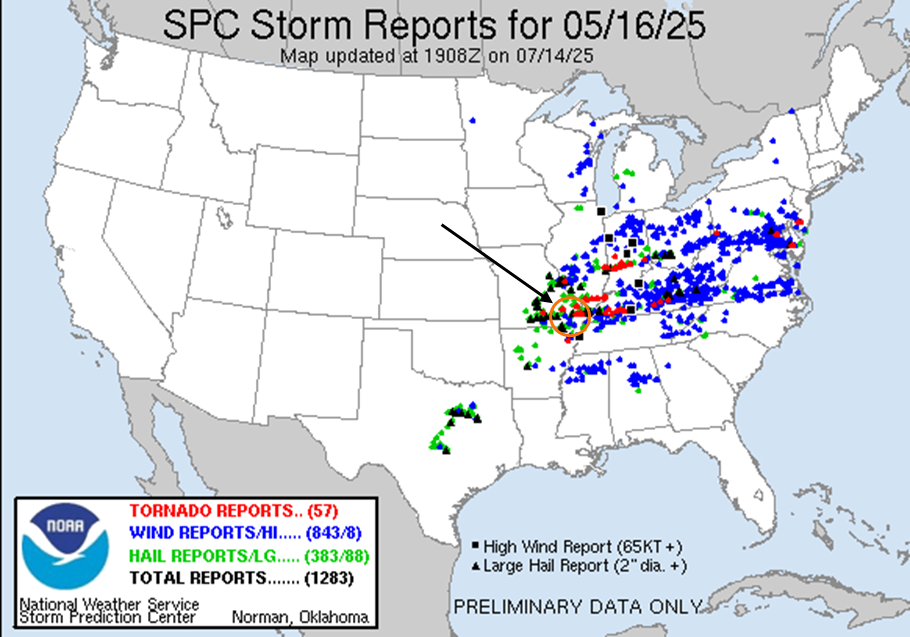}
    \caption{Storm reports from the NOAA Storm Prediction Center for 5/15/25, when the muon detector was located in SE Missouri. The area of muon deployments is circled in orange and pointed to with the arrow.}
    \label{fig:spcreport}
\end{figure}

The detector described above was deployed during late spring/early summer 2025. While the muon detector was not originally designed to collect data while in transit, field modifications were done on 5/14/25 to enable mobile muon flux measurements, albeit without directional capabilities. A successful deployment occurred on 5/16/25 in southeastern Missouri. On this day, there was a geographically expansive severe weather outbreak associated with 49 tornadoes, 180 hail reports, and 538 severe thunderstorm wind reports~\cite{SPCStormReport}. The location of the observations taken in this paper is shown in figure \ref{fig:spcreport}.

In total, muographic data were collected in 3 different thunderstorms. Of these, two were supercells containing mesocyclones, one of which produced a tornado during the observation period. The third deployment was ahead of a non-rotating line of storms that contained neither a mesocyclone nor a tornado. The following sections detail these observations and the corresponding analysis of muon and radar data.

\begin{figure}[h]
    \centering
    \includegraphics[width=0.45\textwidth]{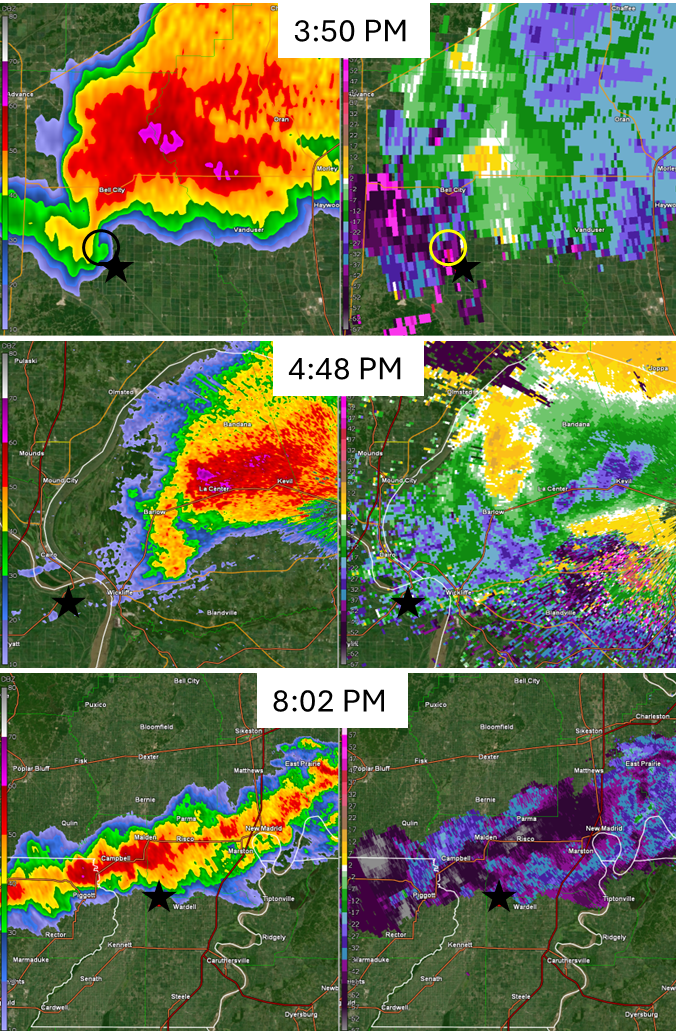}
    \caption{Multi-panel radar imagery from the 3 storm deployments on 5/16/25. Images in the left column are radar reflectivity factor (precipitation distribution and intensity, reds = heavy, blue = light); Right column  images are radial velocity (kts)  indicating motion towards (green, blue, purple) the radar and away (yellow, pink). The pink pixels in the top images are aliased velocities. These were not corrected because the automatic  unfolding algorithm deleted relevant data when dealiasing was applied. The black stars indicate muon detector locations. Scale is equivalent in all 3 images. The circles in the top image indicate the location of the developing tornado . 
}
    \label{fig:radardata}
\end{figure}

\begin{figure}[h!]
    \centering
    \includegraphics[width=0.45\textwidth]{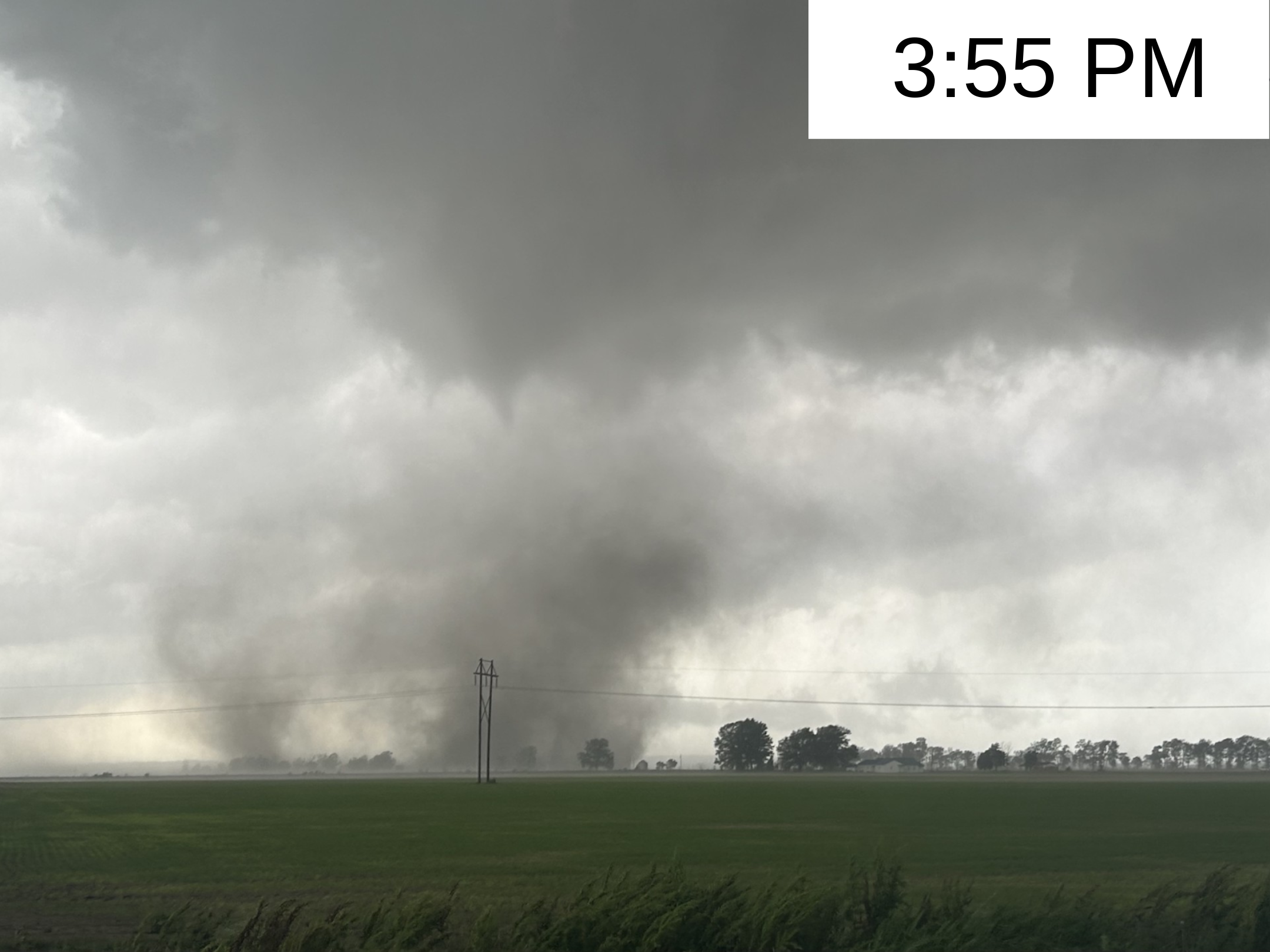}
    \includegraphics[width=0.45\textwidth]{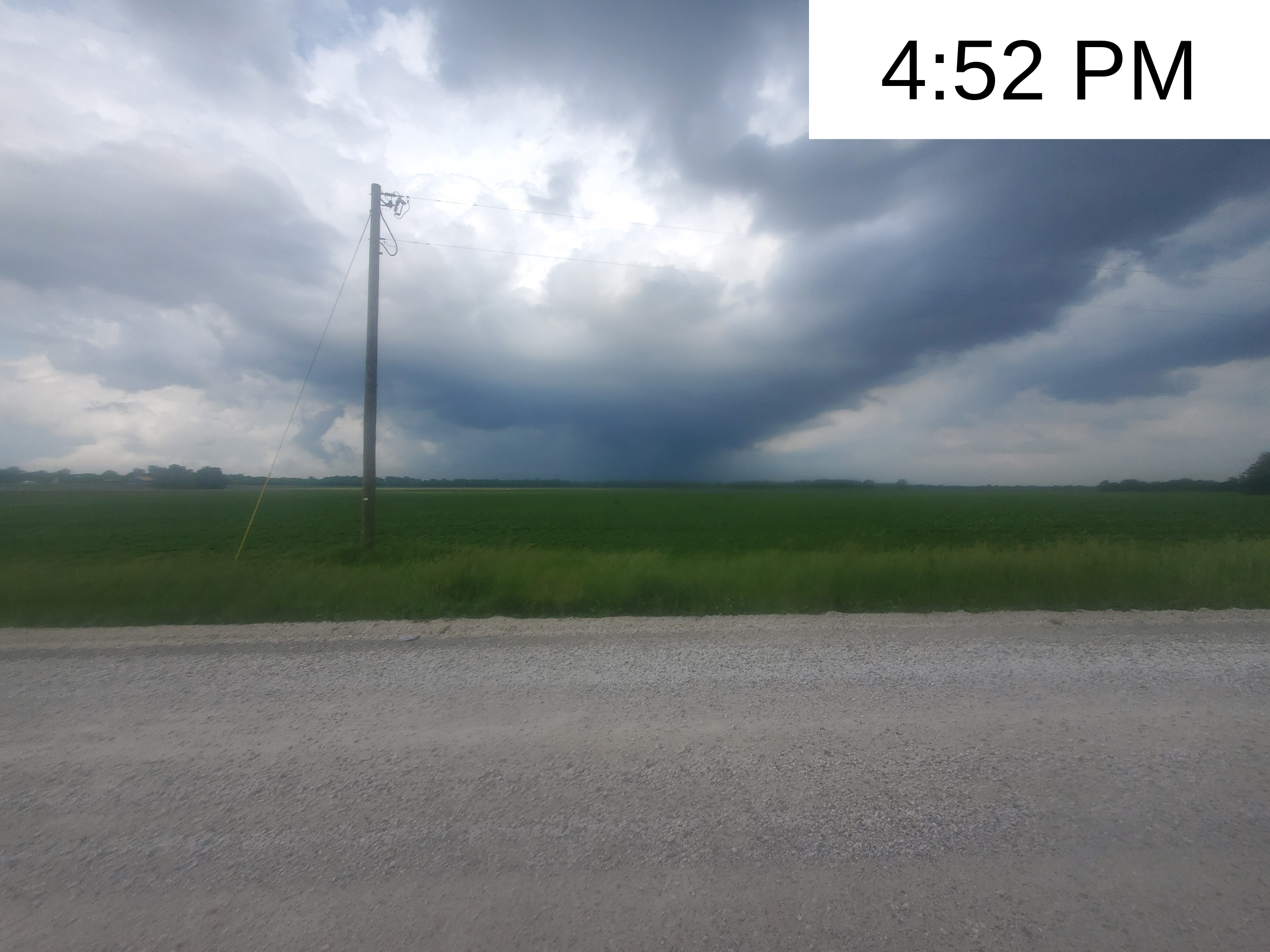}
    \includegraphics[width=0.45\textwidth]{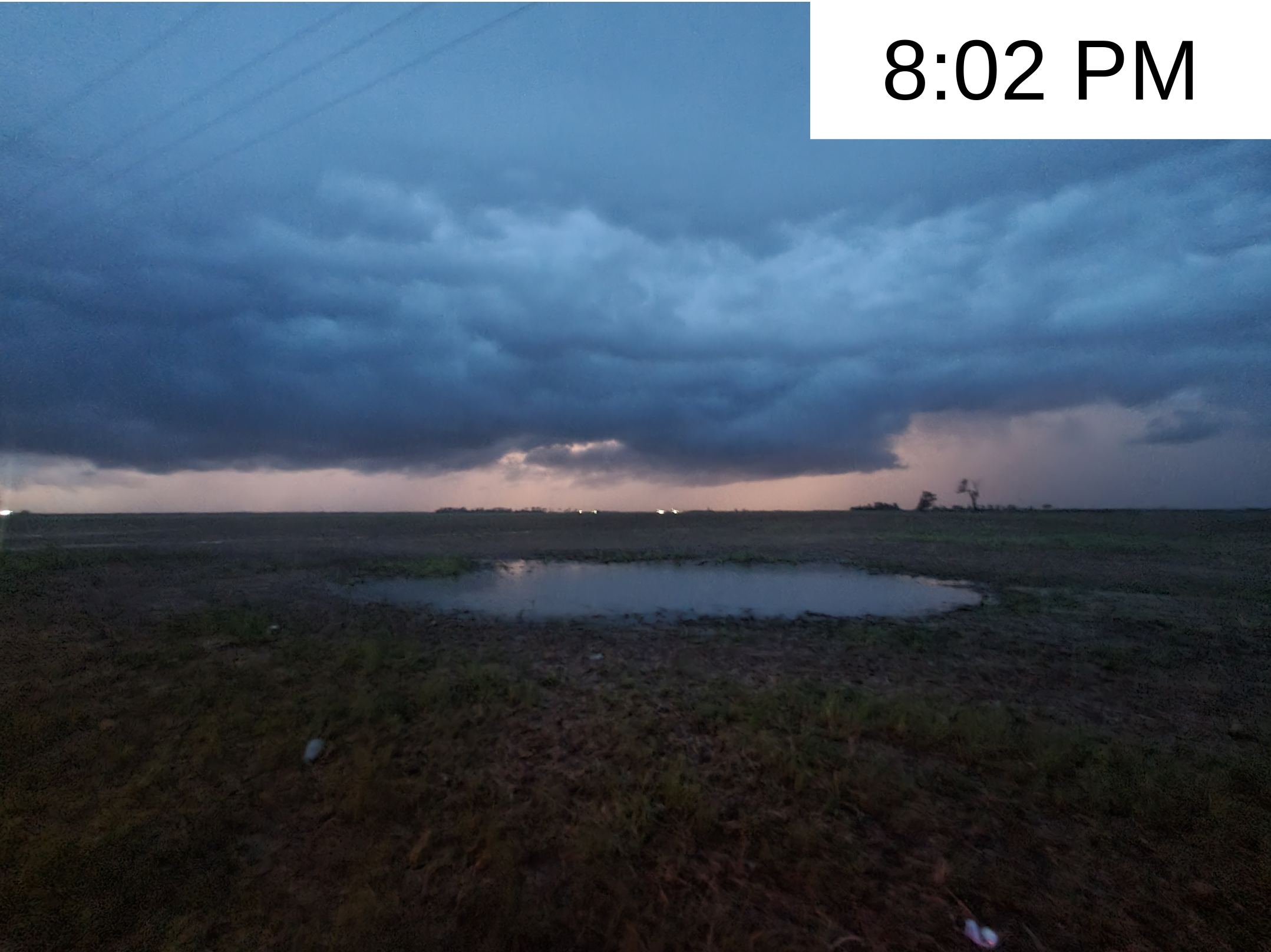}
    \caption{Photos of the observed storms. Top: Photo of the forming tornado, taken just down the street from the muon detector’s observation location on MO N south of Bell City at 3:55 PM CDT. Middle: A photo of the mesocyclone observed during observation A near MO 301 at 4:52 PM CDT, taken from the location of the muon detector. At this point in time, the mesocyclone was $>20$ kilometers from the muon detector location. Bottom: A photo of the non-tornadic line of storms near Gideon, MO at 8:02 PM CDT. This system did not contain a mesocyclone. }
    \label{fig:obs1photo}
\end{figure}

\subsection{Tornado near Blodgett, MO (5/16/25, 3:44 CDT)}

\begin{figure}[h]
    \centering
    \includegraphics[width=0.45\textwidth]{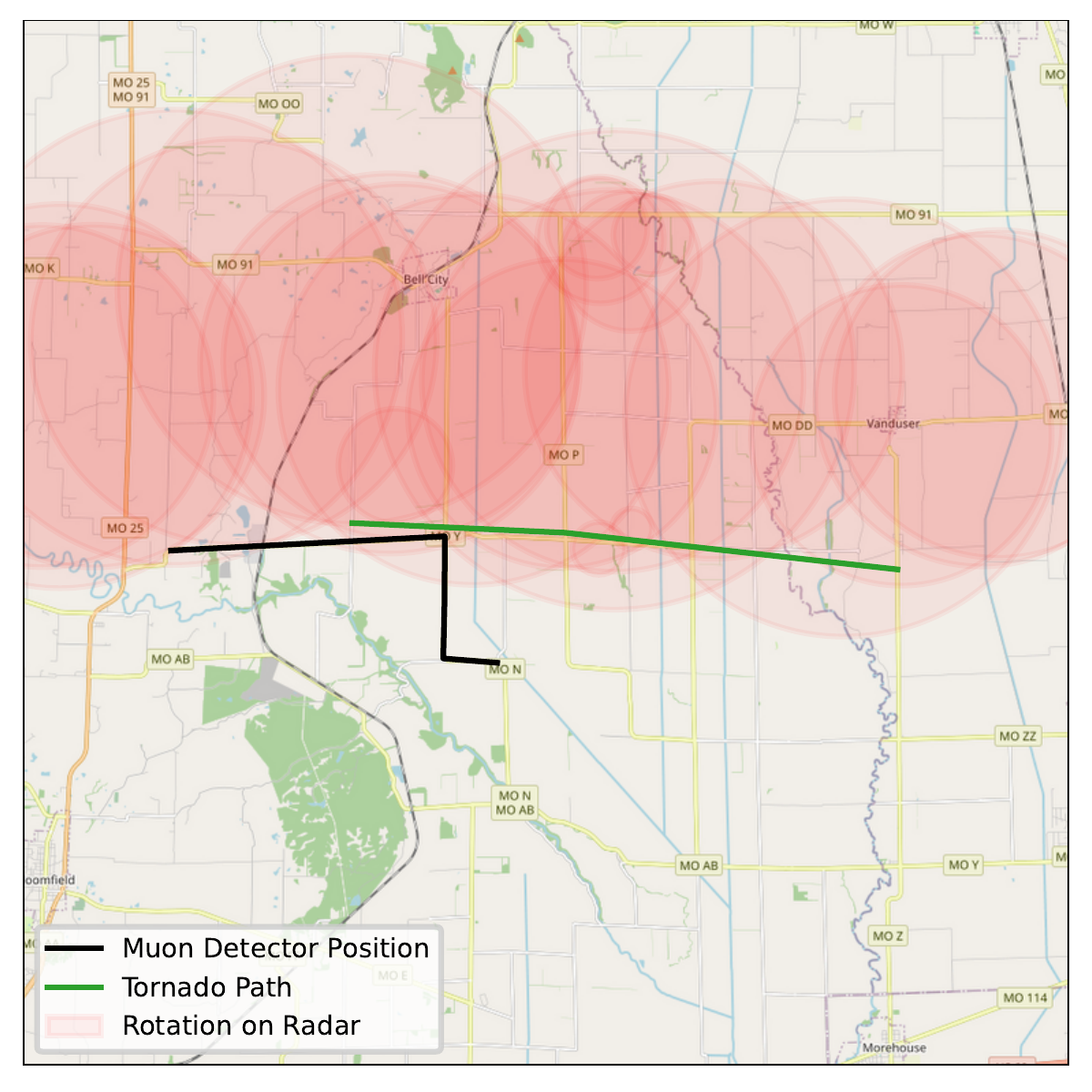}
    \caption{The position of the muon detector (black) and tornado (green) between 3:47 and 3:59 on 5/16/25. Rotation signatures on radar are shown as red circles. }
    \label{fig:mesopos}
\end{figure}

\subsubsection{Meteorological State}

The first deployment of the day occurred between the times of 3:47-3:59 PM CDT (2047-2059 UTC), as the deployment team was positioned south of Bell City, MO, traveling east on MO Y. The team was just to the east of the hook echo (the part of the storm that produces tornadoes) of an intensifying supercell (Figure \ref{fig:radardata}) with a robust elevated mesocyclone (between ~2-6 km above ground level) and an intensifying low-level mesocyclone (in the lowest several 100’s of m above ground level), which was associated with the tornado. At 3:51 PM CDT a tornado began forming behind the deployment vehicle, less than 1 km away. Figure \ref{fig:obs1photo} shows the state of the tornado at 3:55 PM CDT, shortly following its initial formation. The muon detector was located near the trees on the right hand of the image. Figure \ref{fig:mesopos} shows the position of the muon detector and forming tornado relative to the local road system. Rotation signatures from radar obtained from review of radar data (as in figure \ref{fig:radardata}) are shown as red circles. 


\begin{figure}[h]
    \centering
    \includegraphics[width=0.45\textwidth]{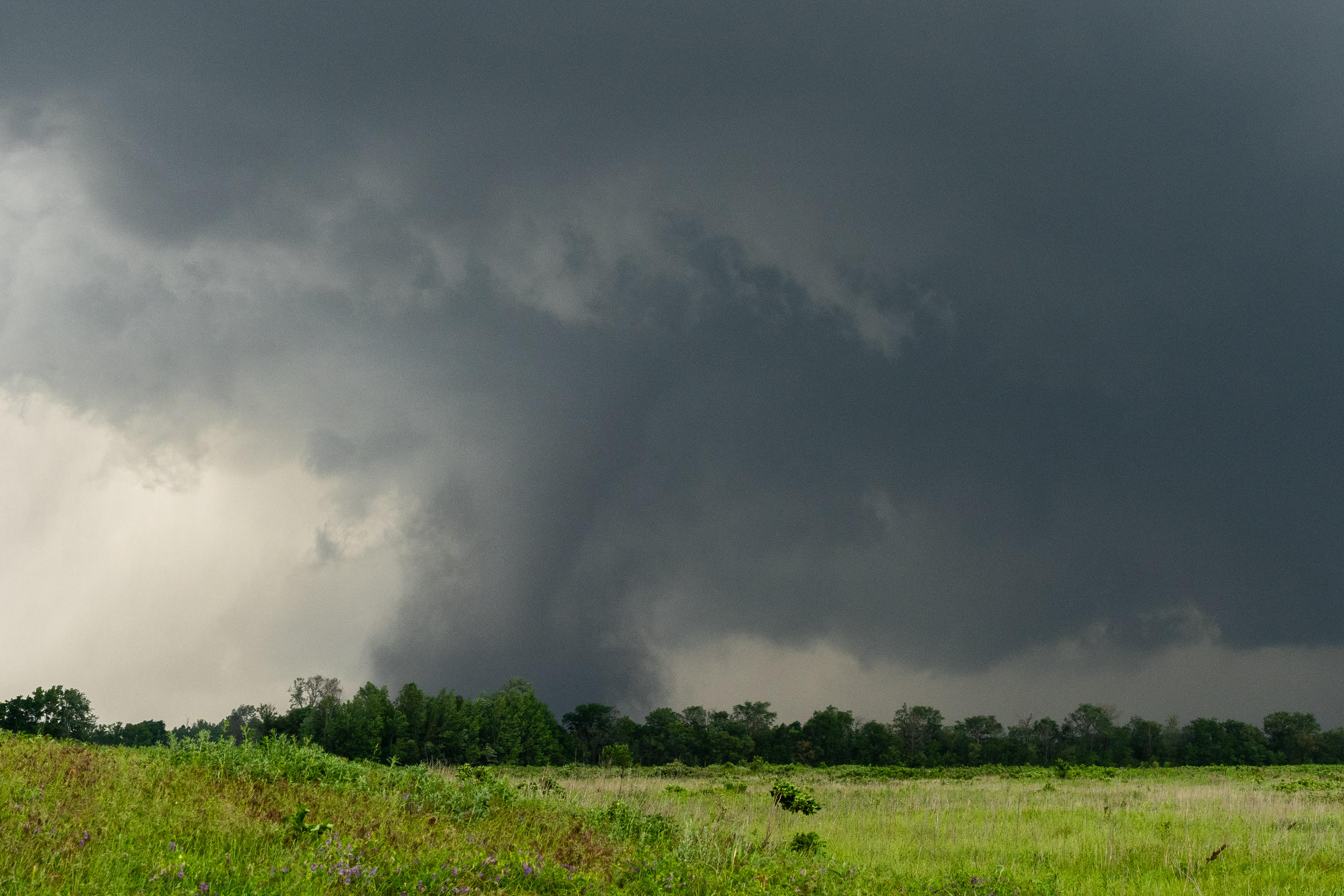}
    \caption{Photo of the tornado at 4:11 PM, after it had strengthened to EF-3 intensity, but left range of the muon detector. Photo coutesey of Greg Robbin\footnote{\url{https://www.gregrobbin.com/}}}
    \label{fig:obs3photo}
\end{figure}

\subsubsection{Muon Observations}

\begin{table*}[t]
    \centering
    \begin{tabular}{| c | c | c | c | c | c |} 
     \hline
       \makecell{\textbf{Time}\\  \textbf{(US Central Time)}} & \makecell{\textbf{Latitude} \\ \textbf{[deg]}} & \makecell{\textbf{Longitude} \\ \textbf{[deg]}} & \makecell{\textbf{Total Muon Rate} \\ \textbf{[Hz]}} & \makecell{\textbf{E-test} \\ \textbf{\textit{p}}} & \makecell{\textbf{Distance to Storm} \\ \textbf{[km]}}\\
     \hline\hline
     \makecell{5:34 - 6:11 PM}  & 36.9211 & -89.3140 & 182.71 $\pm$ 0.29 & (Control Period) & $>15$ \\  
     \hline
     \makecell{3:47 - 3:59 PM}  & 36.9211 & -89.3140 & 183.66 $\pm$ 0.51 & 0.023 & 0 \\  
     \hline
    \end{tabular}
    \caption{Nondirectional muon flux measurements taken during the clear-sky control period (top row) and while near a forming tornado (bottom row). Control period rates shown include the correction factor of 1.003 applied to the control rate to account for non-pressure related daily variations of the atmospheric muon flux. The muon rates measured during the control and tornado periods are compared with the Poisson means test.}
    \label{tab:nondir_table}
\end{table*}

Atmospheric muon data were collected as the detector was in motion, traveling east on road MO Y. Information on the detector roll angle was unfortunately unavailable during this time, as the detector was not designed with automatic roll angle measurement capabilities, and manual measurement was impractical during this deployment period. However, the proximity of this measurement to an active tornado is of potential scientific interest, so we elect to analyze the total ambient muon rate observed, which should be more robust against variations in detector roll angle. 

As we cannot rely on directional information to produce a temporally coincident control measurement for this storm, we must instead use data collected later during the same day to attempt to establish a baseline, clear-sky rate for the atmospheric muon flux. This is somewhat non-ideal, as measurements taken at a later time must be corrected for natural daily variations, and are potentially subject to unpredictable localized geomagnetic and solar activity, at the order of roughly 0.5\% per hour. We attempt to control for regular daily variations using the available data, however acknowledge that this procedure may be imperfect and the existing data may not be sufficient to draw a definitive conclusion as to whether a muon flux perturbation was observed near the forming tornado.

Muon data collected near the forming tornado during the 3:47-3:59 CDT time interval was compared to a baseline rate obtained during a clear-sky control period identified later the same day, between 5:34-6:11 PM CDT. The control period was selected to satisfy the following criteria:

\begin{itemize}
    \item All 3 muon detector panels are operational.
    \item There are no mesocyclones, tornadoes, or other storms within 15 km of the muon detector.
    \item The rate of muon observations is stable over the duration (see below for clarification on how this is determined).
\end{itemize}

\begin{figure}[]
    \centering
    \includegraphics[width=0.45\textwidth]{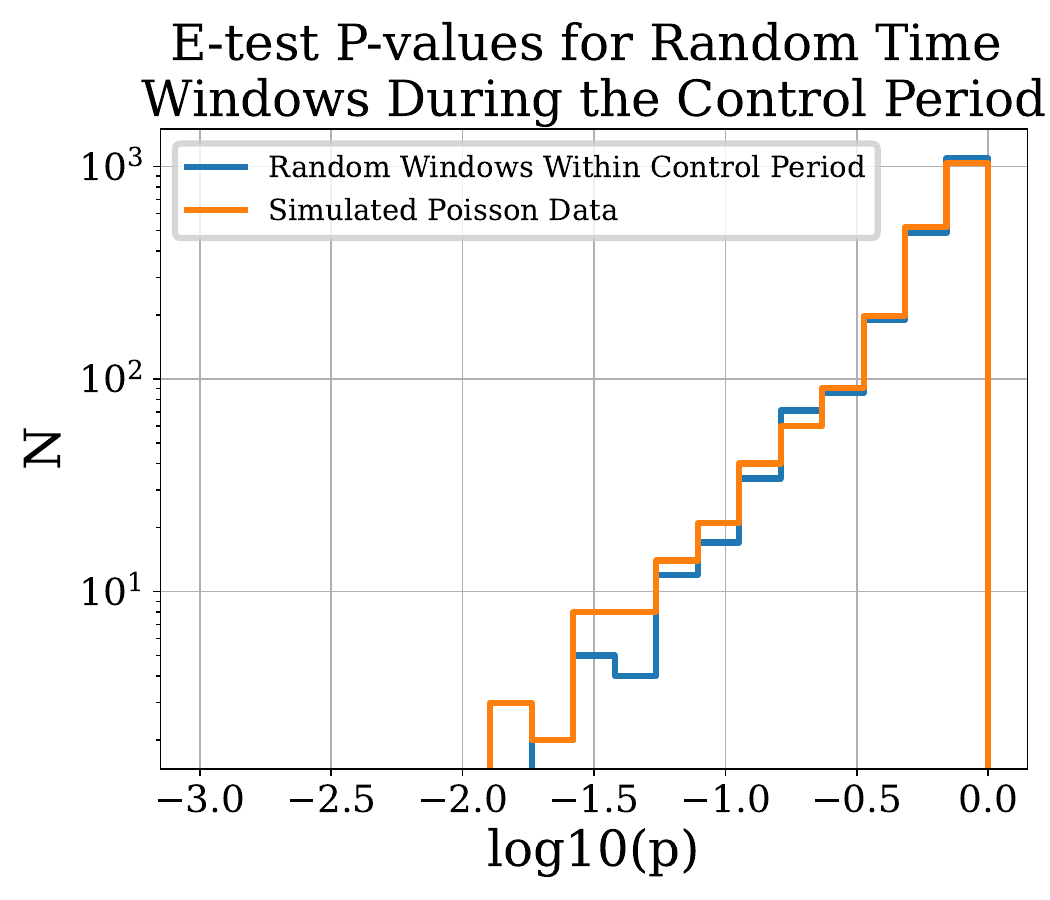}
    \caption{Distributions of the 2 sample E-test p-value when comparing randomly selected time windows within the control period to the overall average control period rate, both for control period data (blue), and simulated data generated according to a poisson distribution (orange). If the muon rate is constant for the duration of the control period, it is expected that the blue and orange distributions should be consistent with one another. A 2-sample KS test comparing these two distributions results in a p-value of 0.31, indicating that the control period data does not display significant rate variations over its duration.}
    \label{fig:controlcheck}
\end{figure}

Stability of control data is determined in two ways:

\begin{enumerate}
    \item The Mann-Kendall test~\cite{mannkendall} is performed for the data binned in 1-second increments over the candidate control measurement duration. This checks for an overall trend of increasing or decreasing muon rates. Data is considered valid for the control period if Mann-Kendall p-value is greater than 0.05. The Mann-Kendall p-value for the selected control period is 0.11.
    \item Random windows within the candidate control measurement period are selected. The rate within these each of these windows is compared to the average rate across the entire control period using the E-test for comparing two measured Poisson rates~\cite{KRISHNAMOORTHY200423} The distribution of resultant p-values should be identical to the distribution generated by performing the same procedure on simulated poisson data. These distributions for the selected control period can be seen in figure \ref{fig:controlcheck}.  
\end{enumerate}

The atmospheric muon rate naturally varies over the duration of the day not only due to weather, but also due solar activity. Long-duration baseline data taken over the course of 3 weeks post-deployment indicate that the atmospheric muon flux between 5:37 and 6:11 CDT on any given day is on average 0.2$\pm 0.1$\% lower than the flux measured between 3:47 and 3:59 CDT. We therefore correct for this effect by adjusting our measured control period rate by a factor of 1.003 when comparing with the on-time data taken near the tornado on 5/16/25.

In principle, measured muon rates should also be controlled for the elevation (i.e. height above sea level) of the detector, as detectors at a higher elevation will naturally measure a higher muon flux. In practice, the deployment region was observed to be impressively flat, with differences in elevation between the control and on-time periods of less than 3 meters, save for a brief duration during the beginning of the on-time period where the detector was traveling through a set of hills south of Bell City between 3:44 and 3:47 CDT. We exclude this period, ensuring to only use data collected while traversing flat ground between 3:47 and 3:59 CDT.

Results from comparing the measured muon flux between 3:47 and 3:59 CDT with the control period between 5:37 and 6:11 CDT can be seen in table \ref{tab:nondir_table}. After correcting for periodic daily variations in the muon flux, a 0.5\% excess of muons was observed when the detector was near a tornado, with a corresponding E test p-value of 0.023 ($2\sigma$). The results of converting the observed muon flux excess to an estimate of the average atmospheric density perturbation within this storm can be seen as the green region in in figure \ref{fig:densityfits}. This estimate appears consistent with projections from existing simulations of tornadic storms~\cite{Orf2019-kn}.

\subsection{Mesocyclone near MO 301 (5/16/25, 4:52 CDT)}
\begin{table*}[t]
    \centering
    \begin{tabular}{| c | c | c | c | c | c | c | c |} 
     \hline
       \makecell{\textbf{Time}\\  \textbf{(US Central Time)}} & \makecell{\textbf{Latitude} \\ \textbf{[deg]}} & \makecell{\textbf{Longitude} \\ \textbf{[deg]}} & \makecell{\textbf{Rate Towards} \\ \textbf{Storm [Hz]}} & \makecell{\textbf{Rate Away From} \\ \textbf{Storm [Hz]}}& \makecell{\textbf{E-test} \\ \textbf{\textit{p}}} & \makecell{\textbf{Distance to Storm} \\ \textbf{[km]}}\\
     \hline\hline
     \makecell{4:52 - 5:11 PM}  & 36.9616 & -89.8917 & 91.06 $\pm$ 0.34 & 90.79 $\pm$ 0.34 & 0.29 & $>16$ \\  
     \hline
     \makecell{8:04 - 8:14 PM}  & 36.3645 & -89.9119 & 87.94 $\pm$ 0.38 & 90.47 $\pm$ 0.39 & $3.6 \times 10^{-6}$ & 4.5 \\  
     \hline
    \end{tabular}
    \caption{Directional muon flux measurements taken near storms on 5/16/25. Rates shown are corrected for detector roll angle, and distances are reported as the distance to the closest edge of the storm estimated from radar data.}
    \label{tab:dir_table}
\end{table*}

\subsubsection{Meteorological State}
The chase team attempted to catch up to the storm for a second deployment after the tornado passed to the east and the team observed it for several minutes. However, due to a combination of non-ideal road options, and a fast forward storm motion, the team was unsuccessful. At 4:52 local time, the decision was made to let the storm continue east into Kentucky and abandon it, in favor of new storms that were approaching from the west. The team parked along MO 301 and deployed the muon detector from 4:52 through 5:11 PM CDT for one final period of data collection. At this time, the supercell had weakened and the tornado had decayed. (Fig \ref{fig:radardata} middle row)  However, there was still an elevated mesocyclone aloft, approximately 16 km from the muon detector. 

\subsubsection{Muon Observations}
For this deployment, the detector was stationary and oriented such that the front of the trailer was facing south, resulting in one side of the detector facing west and the other facing east. The detector roll angle was measured to be 1.5 degrees. Unfortunately, the detector was deployed 16 km away from the target storm, with the storm rapidly moving away from the detector during the data collection period. This makes this observation a weak candidate for measuring the muon flux perturbation associated with this storm. However, due to the unknown magnitude and extent of the density perturbation at a moderate distance from a storm, we choose to analyze this period for the sake of completeness.

Data taken from each direction during this period was adjusted to compensate for the detector roll angle, then compared using the Poisson means test (``E-test")~\cite{KRISHNAMOORTHY200423}. The results can be seen in table \ref{tab:dir_table}. For this observation, no difference in muon flux was observed between the two directions. This appears to be consistent with the initial evaluation of the detector being out of range of the storm system of interest. 

\begin{figure*}[]
    \centering
    \includegraphics[width=0.9\textwidth]{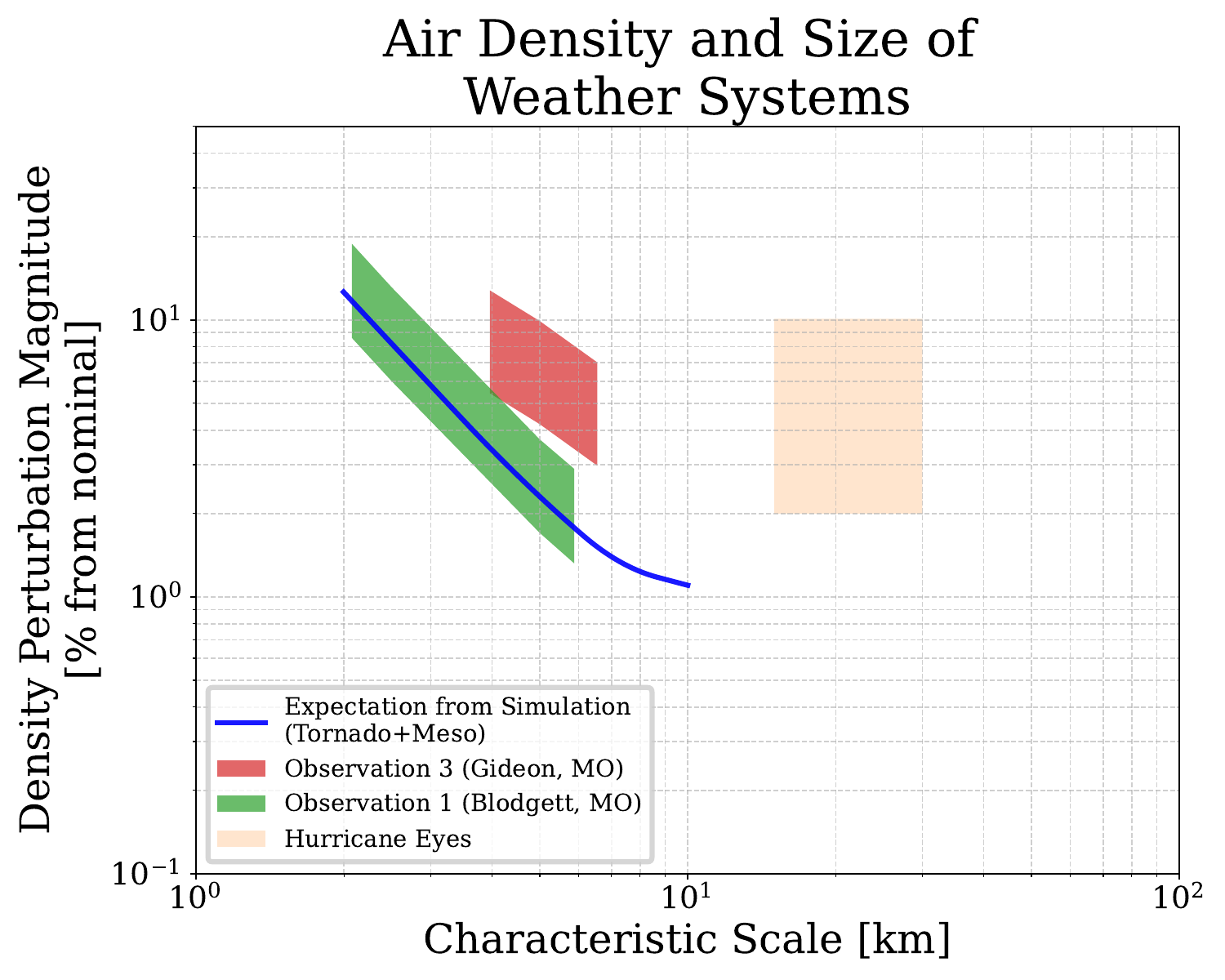}
    \caption{Fits for the size and air density perturbation of storms observed on 5/16/25. Constraints on the characteristic scale of the storms were obtained from radar data, while density perturbation constraints were inferred from muon flux measurements. A range of expectations for the measured density of tornadic storms using simulations from \cite{PhysRevD.111.023018} and \cite{Orf2019-kn} are shown as a blue line spanning the range of characteristic scales for these types of storms. Also shown are estimated regions corresponding to hurricane eyes (orange), which are known to have similarly scaled density perturbations, albeit over a larger volume.}
    \label{fig:densityfits}
\end{figure*}

\subsection{Non-Tornadic Line of Storms near Gideon MO (5/16/25, 8:04 CDT)}
\subsubsection{Meteorological State}
As the convective cells interacted with each other, the storm organizational mode transitioned from discrete supercells to a linear structure. Such upscale growth is common over the course of an afternoon and evening. Initially, the originally discrete supercells maintained their mesocyclones as the line was forming, but eventually, all supercellular characteristics were lost and the line of storm no longer has any rotational elements. While this transition was occurring, the muon detector was deployed for a final time, on the side of road MO EE between 8:04 and 8:14 PM local time. At this time, the line of storms had intense precipitation and weak areas of low-level rotation, but no deep, strong mesocyclones (Fig. \ref{fig:radardata} bottom row).


\subsubsection{Muon Observations}
The detector was stationary and oriented with the front of the trailer facing west, such that one side of the detector was facing a line of storms to the north, and the other was facing clear air to the south. The detector roll angle during this period was measured to be 6.8 degrees.

The results of comparing roll-corrected data from the two opposite directions can be seen in table \ref{tab:dir_table}. Interestingly, the storm appears to have induced a muon deficit at the $p=3.6 \times 10^{-6}$ ($4.5\sigma$) level, rather than an excess, suggesting overall higher density than the surroundings. Meteorologically, this is feasible as linear convective systems often accumulate high density, cold air near the surface through evaporation. The muon detector was within close proximity to the cold air, implying that there could have been a contribution to the mun flux associated with this density perturbation. However, the presence of lightning associated with this storm also suggests that at least a portion of the muon flux deficit could be associated with atmospheric electric fields.

Electric fields within thunderstorms are inconsistent from storm to storm, and the orientation of the electric field vector is not always in the same direction. However, there has been theoretical and experimental work done on understanding the effect on the atmospheric muon flux~\cite{efieldmuons, PhysRevD.89.093005}. The muon flux deficit observed in this paper is roughly a factor of 2 larger than the maximal average deficit observed in \cite{efieldmuons}, suggesting that either the electric fields in the system near Gideon MO were stronger than any of those observed in the previous study, or that there is a non-negligible effect due to local atmospheric density. For the purposes of this paper, we elect to subtract the maximal effect seen in \cite{efieldmuons} from our observed muon flux perturbation magnitude, and use the remainder to obtain a fit for local density within the storm. The results of this calculation can be seen as the red box in figure \ref{fig:densityfits}.

\section{Conclusion}
In this paper, we have presented the results of a field campaign intended to measure the atmospheric muon flux near tornadic storms in the US central plains. Three storms were observed on 5/16/25. Two of these storms produced a measurable effect on the atmospheric muon flux, with the third suspected to have been out of range of the detector during the time of observation.

The measured muon flux perturbations were used to infer the density field perturbation magnitude associated with each of these storms. The results suggest there was an observable low density perturbation (between 1.5 and 11\%, depending on the exact estimate of storm radius) within the tornadic storm and a high density perturbation (between 3 and 11.5\%) in the non-tornadic line of storms. This study marks the first deployment of a muon detector near tornadic storms, indicating that this approach to meteorological measurement is at least logistically possible, however follow-up studies and further development are likely needed to fully address systematic effects such as electric fields other non-weather-related variations in the atmospheric muon flux. 

Notably, the muon detector used in this study was a proof-of-concept device, and future instrumentation designed to be more sensitive could vastly improve measurement capabilities in this context. A larger, tracking muon detector would likely offer significantly improved muon flux sensitivity in addition to allowing for directional localization of measured density field features, rather than simply an average over the entire storm. Improvements in the understanding of electric fields within tornadic storms, and/or dedicated instrumentation to measure the local electric field near the deployment location would be similarly helpful, as this would reduce the main source of uncontrolled uncertainty for this type of measurement.

\section{Acknowledgments}
This work was supported by the Ohio State President's Research Excellence Accelerator award. The authors would like to thank John Beacom, Paul Martini, Peter Taylor, and Austin Cummings for helpful discussion and support for the duration of this project. The authors would also like to thank Greg Robbin, and the students of ATMOSSC 5701 for sharing photos and videos of the storms observed. 

\clearpage
\bibliography{apssamp}

@article{RadiationTextbook,
author = {Sigmund, Peter},
year = {2008},
month = {01},
pages = {151},
title = {Particle Penetration and Radiation Effects},
journal = {Particle Penetration and Radiation Effects, by Peter Sigmund. Berlin: Springer, 2008. ISBN: 978-3-540-72622-7}
}

@article{LECHMANN2021103842,
title = {Muon tomography in geoscientific research – A guide to best practice},
journal = {Earth-Science Reviews},
volume = {222},
pages = {103842},
year = {2021},
issn = {0012-8252},
doi = {https://doi.org/10.1016/j.earscirev.2021.103842},
url = {https://www.sciencedirect.com/science/article/pii/S0012825221003433},
author = {Alessandro Lechmann and David Mair and Akitaka Ariga and Tomoko Ariga and Antonio Ereditato and Ryuichi Nishiyama and Ciro Pistillo and Paola Scampoli and Fritz Schlunegger and Mykhailo Vladymyrov}
}

@misc{tilav2019seasonal,
      title={Seasonal variation of atmospheric muons in IceCube}, 
      author={Serap Tilav and Thomas K. Gaisser and Dennis Soldin and Paolo Desiati},
      year={2019},
      eprint={1909.01406},
      archivePrefix={arXiv},
      primaryClass={astro-ph.HE}
}

@article{Jourde_2016,
   title={Monitoring temporal opacity fluctuations of large structures with muon radiography: a calibration experiment using a water tower},
   volume={6},
   ISSN={2045-2322},
   url={http://dx.doi.org/10.1038/srep23054},
   DOI={10.1038/srep23054},
   number={1},
   journal={Scientific Reports},
   publisher={Springer Science and Business Media LLC},
   author={Jourde, Kevin and Gibert, Dominique and Marteau, Jacques and de Bremond d’Ars, Jean and Gardien, Serge and Girerd, Claude and Ianigro, Jean-Christophe},
   year={2016},
   month=mar }

@article {PMID:31040358,
	Title = {First muography of Stromboli volcano},
	Author = {Tioukov, Valeri and Alexandrov, Andrey and Bozza, Cristiano and Consiglio, Lucia and D'Ambrosio, Nicola and De Lellis, Giovanni and De Sio, Chiara and Giudicepietro, Flora and Macedonio, Giovanni and Miyamoto, Seigo and Nishiyama, Ryuichi and Orazi, Massimo and Peluso, Rosario and Sheshukov, Andrey and Sirignano, Chiara and Stellacci, Simona Maria and Strolin, Paolo and Tanaka, Hiroyuki K M},
	DOI = {10.1038/s41598-019-43131-8},
	Number = {1},
	Volume = {9},
	Month = {April},
	Year = {2019},
	Journal = {Scientific reports},
	ISSN = {2045-2322},
	Pages = {6695},
	URL = {https://europepmc.org/articles/PMC6491474},
}

@article{typhoons,
author = {Tanaka, Hiroyuki and Gluyas, Jon and Holma, Marko and Joutsenvaara, J. and Kuusiniemi, Pasi and Leone, Giovanni and Lo Presti, D. and Matsushima, Jun and Oláh, László and Steigerwald, Sara and Thompson, Lee and Usoskin, Ilya and Poluianov, Stepan and Varga, Dezső and Yokota, Yusuke},
year = {2022},
month = {10},
pages = {16710},
title = {Atmospheric muography for imaging and monitoring tropic cyclones},
volume = {12},
journal = {Scientific Reports},
doi = {10.1038/s41598-022-20039-4}
}

@article{muthunderstorms2,
title = {Studies of Thunderstorm Events Based on the Data of Muon Hodoscope URAGAN and Meteorological Radar DMRL-C},
journal = {Physics Procedia},
volume = {74},
pages = {486-492},
year = {2015},
note = {Fundamental Research in Particle Physics and Cosmophysics},
issn = {1875-3892},
doi = {https://doi.org/10.1016/j.phpro.2015.09.239},
url = {https://www.sciencedirect.com/science/article/pii/S1875389215014388},
author = {A.V. Kozyrev and N.S. Barbashina and T.A. Belyakova and J.B. Pavlyukov and A.A. Petrukhin and N.I. Serebryannik and V.V. Shutenko and I.I. Yashin}
}

@article{PhysRevD.111.023018,
  title = {Effect of tornadic supercell thunderstorms on the atmospheric muon flux},
  author = {Luszczak, William and Orf, Leigh},
  journal = {Phys. Rev. D},
  volume = {111},
  issue = {2},
  pages = {023018},
  numpages = {12},
  year = {2025},
  month = {Jan},
  publisher = {American Physical Society},
  doi = {10.1103/PhysRevD.111.023018},
  url = {https://link.aps.org/doi/10.1103/PhysRevD.111.023018}
}

@BOOK{GaisserCR,
       author = {{Gaisser}, Thomas K.},
        title = "{Cosmic rays and particle physics.}",
         year = 1990,
       adsurl = {https://ui.adsabs.harvard.edu/abs/1990cup..book.....G}
}

@ARTICLE{Orf2019-kn,
  title     = "A Violently Tornadic Supercell Thunderstorm Simulation Spanning
               a {Quarter-Trillion} Grid Volumes: Computational Challenges,
               {I/O} Framework, and Visualizations of Tornadogenesis",
  author    = "Orf, Leigh",
  journal   = "Atmosphere",
  publisher = "Multidisciplinary Digital Publishing Institute",
  volume    =  10,
  number    =  10,
  pages     = "578",
  month     =  sep,
  year      =  2019,
  doi       = "10.3390/atmos10100578"
}

@article{Kauer:2019len,
    author = "Kauer, Matt and Huber, Thomas and Tosi, Delia and Wendt, Chris",
    collaboration = "IceCube",
    title = "{The Scintillator Upgrade of IceTop: Performance of the prototype array}",
    eprint = "1908.09860",
    archivePrefix = "arXiv",
    primaryClass = "astro-ph.HE",
    reportNumber = "PoS-ICRC2019-309",
    doi = "10.22323/1.358.0309",
    journal = "PoS",
    volume = "ICRC2019",
    pages = "309",
    year = "2021"
}

@article{Beznosko:2005ba,
    author = "Beznosko, D. and Bross, A. and Dyshkant, A. and Pla-Dalmau, A. and Rykalin, V.",
    title = "{FNAL-NICADD extruded scintillator}",
    journal = "FERMILAB-PUB-05-344",
    month = "9",
    year = "2005",
    URL = {https://lss.fnal.gov/archive/2005/pub/fermilab-pub-05-344.pdf}
}

@article{efieldmuons,
author = {Chilingarian, A. and Hovsepyan, G. and Zazyan, M.},
title = {Muon Tomography of Charged Structures in the Atmospheric Electric Field},
journal = {Geophysical Research Letters},
volume = {48},
number = {17},
pages = {e2021GL094594},
doi = {https://doi.org/10.1029/2021GL094594},
url = {https://agupubs.onlinelibrary.wiley.com/doi/abs/10.1029/2021GL094594},
eprint = {https://agupubs.onlinelibrary.wiley.com/doi/pdf/10.1029/2021GL094594},
note = {e2021GL094594 2021GL094594},
year = {2021}
}

@article{PhysRevD.89.093005,
  title = {Variations of muon flux in the atmosphere during thunderstorms},
  author = {Karapetyan, G. G.},
  journal = {Phys. Rev. D},
  volume = {89},
  issue = {9},
  pages = {093005},
  numpages = {8},
  year = {2014},
  month = {May},
  publisher = {American Physical Society},
  doi = {10.1103/PhysRevD.89.093005},
  url = {https://link.aps.org/doi/10.1103/PhysRevD.89.093005}
}

@article { AStudyoftheTornadicRegionwithinaSupercellThunderstorm,
      author = "Joseph B.  Klemp and Richard  Rotunno",
      title = "A Study of the Tornadic Region within a Supercell Thunderstorm",
      journal = "Journal of Atmospheric Sciences",
      year = "1983",
      publisher = "American Meteorological Society",
      address = "Boston MA, USA",
      volume = "40",
      number = "2",
      doi = "10.1175/1520-0469(1983)040<0359:ASOTTR>2.0.CO;2",
      pages=      "359 - 377",
      url = "https://journals.ametsoc.org/view/journals/atsc/40/2/1520-0469_1983_040_0359_asottr_2_0_co_2.xml"
}

@article { BuoyancyandPressurePerturbationsDerivedfromDualDopplerRadarObservationsofthePlanetaryBoundaryLayerApplicationsforMatchingModelswithObservations,
      author = "Tzvi  Gal-Chen and Robert A.  Kropfli",
      title = "Buoyancy and Pressure Perturbations Derived from Dual-Doppler Radar Observations of the Planetary Boundary Layer: Applications for Matching Models with Observations",
      journal = "Journal of Atmospheric Sciences",
      year = "1984",
      publisher = "American Meteorological Society",
      address = "Boston MA, USA",
      volume = "41",
      number = "20",
      doi = "10.1175/1520-0469(1984)041<3007:BAPPDF>2.0.CO;2",
      pages=      "3007 - 3020",
      url = "https://journals.ametsoc.org/view/journals/atsc/41/20/1520-0469_1984_041_3007_bappdf_2_0_co_2.xml"
}

@article { PerturbationPressureFieldsMeasuredbyAircraftaroundtheCloudBaseUpdraftofDeepConvectiveClouds,
      author = "Margaret A.  LeMone and Gary M.  Barnes and James C.  Fankhauser and Lesley F.  Tarleton",
      title = "Perturbation Pressure Fields Measured by Aircraft around the Cloud-Base Updraft of Deep Convective Clouds",
      journal = "Monthly Weather Review",
      year = "1988",
      publisher = "American Meteorological Society",
      address = "Boston MA, USA",
      volume = "116",
      number = "2",
      doi = "10.1175/1520-0493(1988)116<0313:PPFMBA>2.0.CO;2",
      pages=      "313 - 327",
      url = "https://journals.ametsoc.org/view/journals/mwre/116/2/1520-0493_1988_116_0313_ppfmba_2_0_co_2.xml"
}

@misc{SPCStormReport,
  author = {{National Weather Service, Storm Prediction Center}},
  title = {Overview of the Tornado Outbreak of May 16, 2025},
  howpublished = {\url{https://www.weather.gov/pah/2025May16_Severe}},
  note = {Accessed: 20 January 2026},
  year = {2025},
  month = {May},
  day = {16}
}

@article{mannkendall,
author = {Henry B. Mann},
title = {Nonparametric Tests Against Trend},
journal = {Econometrica},
volume = {13},
number = {3},
pages = {245-259},
url = {https://onlinelibrary.wiley.com/doi/abs/0012-9682(194507)13:3&lt;245:NTAT&gt;2.0.CO;2-U},
year = {1945}
}

@article{KRISHNAMOORTHY200423,
title = {A more powerful test for comparing two Poisson means},
journal = {Journal of Statistical Planning and Inference},
volume = {119},
number = {1},
pages = {23-35},
year = {2004},
issn = {0378-3758},
doi = {https://doi.org/10.1016/S0378-3758(02)00408-1},
url = {https://www.sciencedirect.com/science/article/pii/S0378375802004081},
author = {K. Krishnamoorthy and Jessica Thomson}
}

\end{document}